\documentclass[12pt]{article}
\usepackage{graphicx,amsmath}
\usepackage{units}

\newlength{\dinwidth}
\newlength{\dinmargin}
\renewcommand{\vec}[1]{\boldsymbol{#1}}

\def\lapproxeq{\lower .7ex\hbox{$\;\stackrel{\textstyle                                                    
<}{\sim}\;$}}                                                    
\def\gapproxeq{\lower .7ex\hbox{$\;\stackrel{\textstyle                                                    
>}{\sim}\;$}}                                                    
\def\be{\begin{equation}}                                                    
\def\ee{\end{equation}}                                                    
\def\bea{\begin{eqnarray}}                                                    
\def\eea{\end{eqnarray}}
\def\b{\vec{b}}
     
\def\q{\vec{q}}

\def\GeV{\rm GeV}

\def\sh{\hat s}
\def\sh2{{\hat s}^2}
\begin{document}
                                                    
\titlepage                                                    
\begin{flushright}                                                    
IPPP/12/04  \\
DCPT/12/08 \\                                                    
\today \\                                                    
\end{flushright} 
\vspace*{0.5cm}
\begin{center}                                                    
{\Large \bf Proton opacity in the light of LHC diffractive data}\\

\vspace*{1cm}
                                                   
M.G. Ryskin$^{a,b}$, A.D. Martin$^a$ and V.A. Khoze$^{a,b}$ \\                                                    
                                                   
\vspace*{0.5cm}                                                    
$^a$ Institute for Particle Physics Phenomenology, University of Durham, Durham, DH1 3LE \\                                                   
$^b$ Petersburg Nuclear Physics Institute, NRC Kurchatov Institute, Gatchina, St.~Petersburg, 188300, Russia

\vspace*{1cm}                                                    
                                                    
\begin{abstract}                                                    
We show that collider data on elastic $pp$ (and $p\bar{p}$) scattering, including the LHC TOTEM data at 7 TeV, can be well described by a 3-channel eikonal model with only one Pomeron, with parameters that are naturally linked to the perturbative QCD (BFKL) framework. The proton opacity, determined in this way, is then used to account for sizeable absorptive effects. We study the  recent measurements of $d\sigma/d\Delta\eta$ made by the ATLAS collaboration, where they select events with large rapidity gaps $\Delta\eta$. We demonstrate that the absorptive corrections noticeably change both the value and the $\Delta\eta$ dependence of the cross section. We find that our parameter-free calculation is in agreement with these ATLAS data. 

\end{abstract}                                                        
\vspace*{0.5cm}                                                    
                                                    
\end{center}                                                    
                                                   
\section {Introduction}

Recent measurements of diffractive processes at the LHC can be used to greatly improve our knowledge of the opacity (the hadronic matter density) of the proton.  In particular, here we study the implications of the TOTEM data \cite{TOTEM1,TOTEM2} and the determinations of the rapidity gap cross sections measured by ATLAS 
at $\sqrt{s}=7 ~{\rm TeV}$ \cite{Atlas}.

It is convenient\footnote{See, for instance, Ref. \cite{abk}.} 
to describe the $pp$ elastic scattering amplitude in the impact parameter, $b$, representation, since at high energy $b$ is well determined. Essentially $b$ is equivalent to the partial wave $\ell=b\sqrt{s}/2$.  It is known that the real part of the high-energy elastic amplitude is small in comparison with the imaginary part. Thus, to good accuracy, the amplitude can be written in the eikonal form
\begin{equation}
T(b)~=~i(1-e^{-\Omega (b) /2}),
\label{eq:1}
\end{equation}
where $\Omega$ is called the opacity or optical density of the proton. The simplest, popular, parametrization describes the opacity by one-Pomeron exchange.  On the other hand, we can determine $T(b)$, and hence the opacity $\Omega (b)$, directly from the elastic data \cite{Amaldi}
\begin{equation}
{\rm Im}T(b)~=~\int \sqrt{\frac{d\sigma_{\rm el}}{dt}\frac{16\pi}{1+\rho^2}}~ J_0(q_tb)~ \frac{q_tdq_t}{4\pi},
\label{eq:2}
\end{equation}
where $q_t=\sqrt{|t|}$ and $\rho \equiv {\rm Re}T/{\rm Im}T$. In this way, we first determine Im$T$ from the data for $d\sigma_{\rm el}/dt$, and then calculate $\Omega (b)$ using {(\ref{eq:1}), assuming $\rho(t)=$constant. In fact, later on, we assume $\rho^2\ll 1$. The results are shown in Fig.~\ref{fig:tt1}, where we compare $\Omega(b)$ obtained from elastic differential cross section data at S$p\bar{p}$S \cite{SppS}, Tevatron \cite{Tev} and LHC \cite{TOTEM2} energies. At the lower two energies the $\Omega (b)$ distributions have approximately Gaussian form, whereas at the LHC energy we observe a growth of $\Omega$ at small $b$. The growth reflects the fact that the TOTEM data indicate that we have almost total saturation at $b=0$. Note that according to (\ref{eq:1}) the
 value of Im$T(b=0)\to 1$ corresponds to $\Omega \to \infty$. 
Since actually we do not reach exact saturation the proton opacity at $b=0$ is not $\infty$, but just numerically large.  Clearly, in this region of $b$ the uncertainty on the value of $\Omega$ is large as well, see Fig.~\ref{fig:tt1}.   
\begin{figure} [htb]
\begin{center}
\includegraphics[height=12cm]{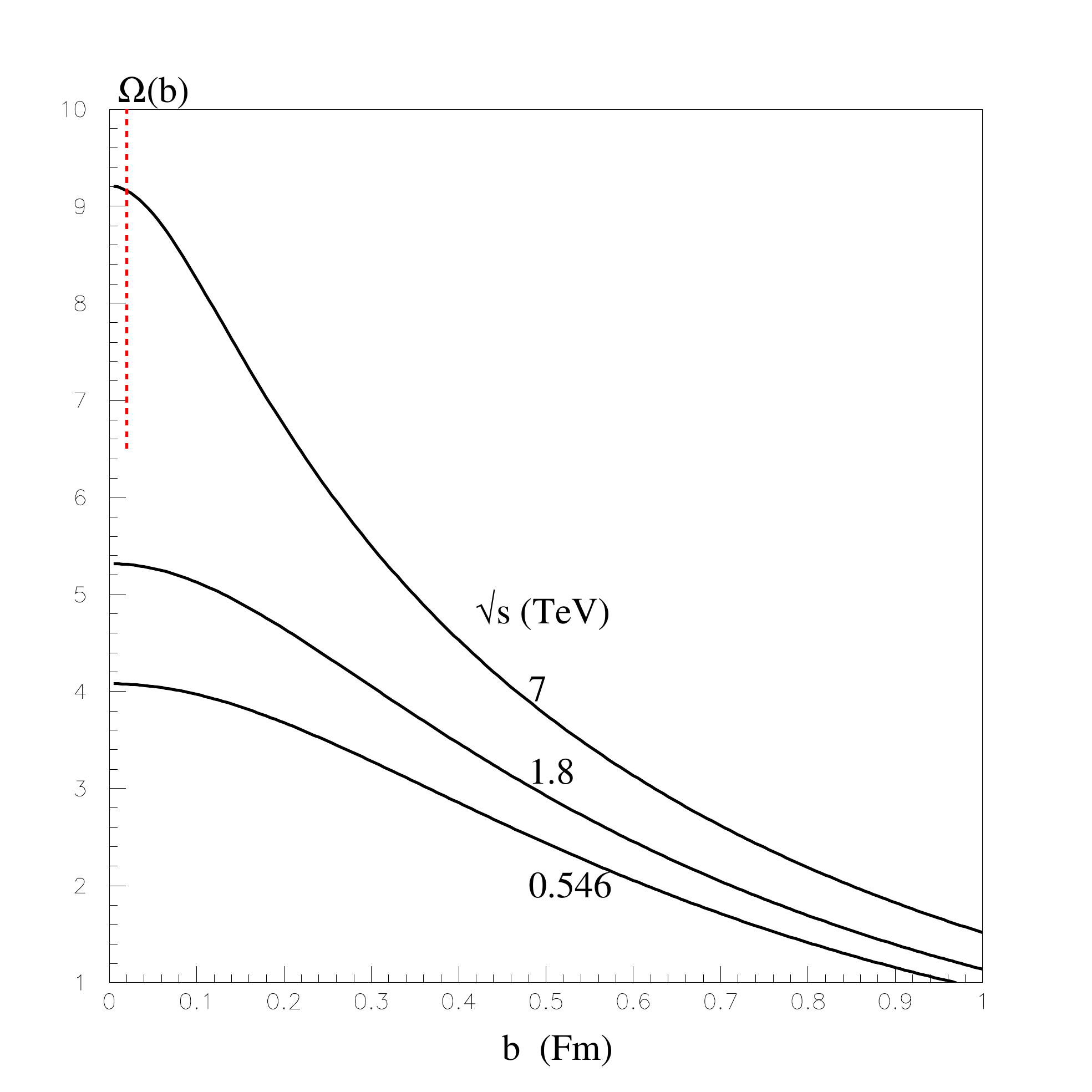}
\vspace{-0.2cm}
\caption{\sf The proton opacity $\Omega(b)$ determined directly from the $pp$ $d\sigma_{\rm el}/dt$ data at 546 GeV \cite{SppS}, 1.8 TeV \cite{Tev} and 7 TeV \cite{TOTEM2} data.  The uncertainty on the LHC value at $b=0$ is indicated by a dashed line.}
\label{fig:tt1}
\end{center}
\end{figure}

One obvious possibility to account for the growth of $\Omega$ is to 
conclude that we observe, at low $b$, a component due to a qualitatively new interaction with an unusually strong energy behaviour. In particular, in Ref.\cite{DL2} the elastic total cross section was described by two Pomerons; one with intercept 0.09, and a second with a much higher intercept $0.36$.  The residue of the latter Pomeron has a relatively flat $t$-behaviour, so that it contributes mainly to the low $b$ region.

However, this is not the only possibility. Recall that the proton is not a local object, but has its own complicated structure. Proton interactions can be mediated via its excited states. An economical way to allow for the effect of the excitations is to use the Good-Walker formalism \cite{GW} of diffractive eigenstates, each of which undergoes ``elastic scattering'' only.  We show below that the present data can be described by a 3-channel eikonal model with 3 diffractive eigenstates of different size, but with only the one Pomeron.

Besides the elastic data of the TOTEM experiment, the other relevant LHC data are the rapidity gap cross section measurements of ATLAS \cite{Atlas}, which are shown\footnote{Actually $\Delta\eta$ may not be the full gap size, but is the part of it that is observed by the calorimeter. Here we have plotted the distribution in terms of $\Delta\eta$. Our $\Delta\eta$ is denoted as $\Delta\eta_F$ in \cite{Atlas}. ATLAS add the subscript $F$ to emphasize that the gap is observed to start at the edge of the calorimeter. Thus the true gap size may be larger than the observed $\Delta\eta$ (or $\Delta\eta_F$ in the notation of \cite{Atlas}).} 
in Fig.~\ref{fig:Atlas}. For a sufficiently large rapidity gap, $\Delta\eta$, these data correspond to high-mass diffractive dissociation, and should be well described in a triple-Pomeron framework in which the rapidity gap survival probability, $S^2$, is taken into account. We emphasis that the region of $b$ sampled in the interaction depends on the mass of the diffractive state, or, equivalently, on the size of the rapidity gap.  As we shall explain in Section \ref{sec:shape}, a gap of larger size corresponds to smaller $b$. On the other hand, at lower $b$ we have a larger opacity $\Omega$, that is a smaller probability of gap survival. We will study the role of the gap survival factor using the 3-channel eikonal model we obtained to describe the TOTEM elastic data.
\begin{figure} [htb]
\begin{center}
\includegraphics[height=10cm]{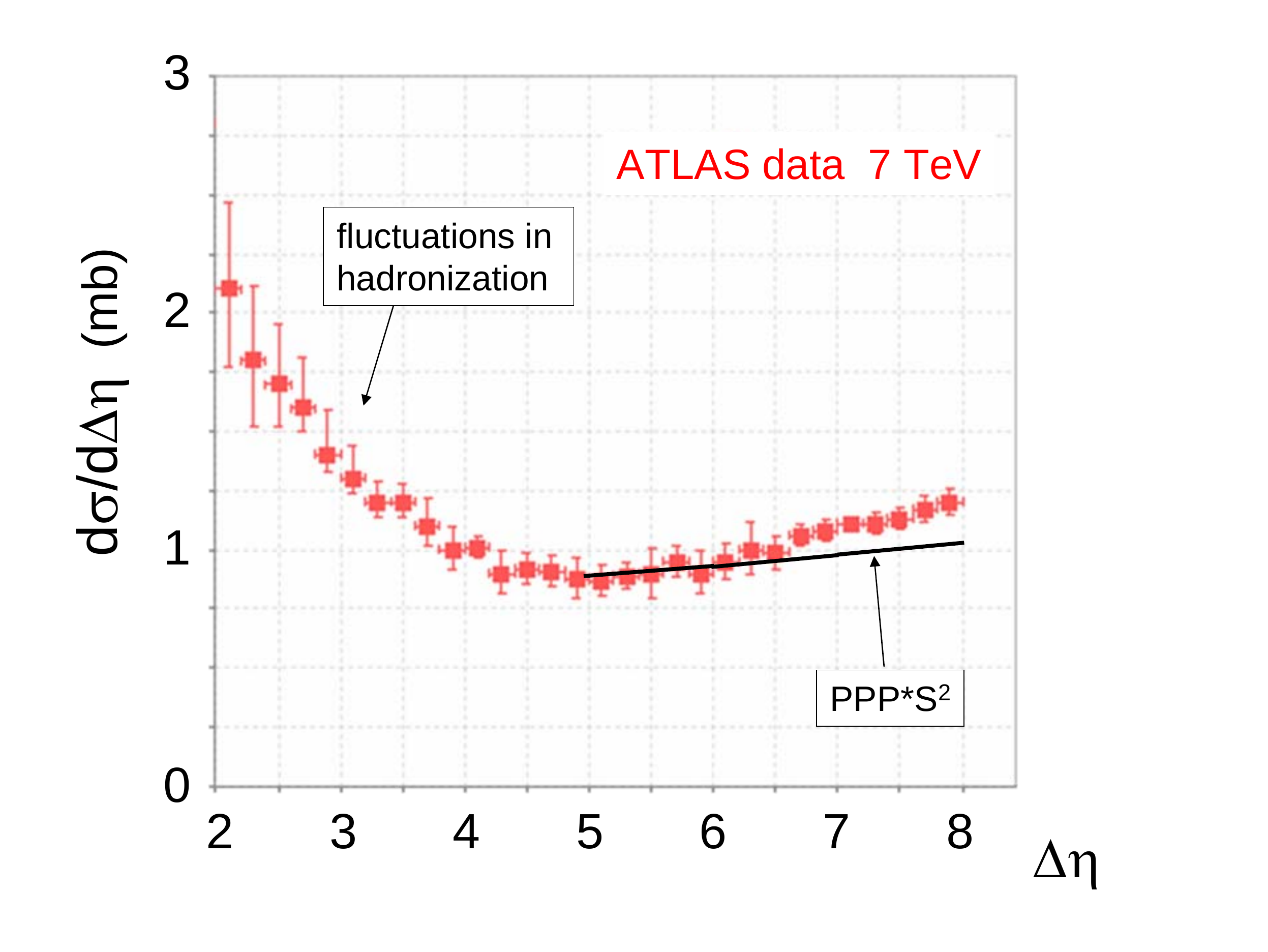}
\vspace{-0.8cm}
\caption{\sf The ATLAS measurements of the inelastic cross section differential in rapidity gap size $\Delta\eta$ for particles with $p_T>200$ MeV \cite{Atlas}. Events with small gap size ($\Delta\eta \lapproxeq 5$) may have a non-diffractive component which arises from fluctuations in the hadronization process \cite{KKMRZ}. This component increases as $\Delta\eta$ decreases (or if a larger $p_T$ cut is used \cite{KKMRZ,Atlas}).  
The data with $\Delta\eta \gapproxeq 5$ are dominantly of diffractive origin, and may be compared with model predictions which allow for the survival probability, $S^2$, of the rapidity gap. The curve is the prediction obtained as described in Section \ref{sec:diff}.}
\label{fig:Atlas}
\end{center}
\end{figure}

\section{Description of elastic high-energy $pp$ scattering  \label{sec:elastic}}

Here we show that it is possible to describe the elastic cross section at 7 TeV in terms of a single Pomeron pole using a multi-channel eikonal.  Indeed, such a formalism is quite natural. It is a consequence of the internal structure of the proton. At high energies the lifetimes of the fluctuations of a fast proton are large, $\tau \sim E/m^2$. During these time intervals the corresponding Fock states can be considered as `frozen'. Each hadronic constituent can undergo scattering and thus destroy the coherence of the fluctuations. As a consequence, the outgoing superposition of states will be different from the incident proton and will, in terms of hadrons, contain not only the proton, but excited proton states as well.

The appropriate formalism was proposed long ago by Good and Walker \cite{GW}. We introduce so-called diffractive eigenstates, $|\phi_i\rangle$ with $i=1,n$, that diagonalize the $T$-matrix, and so only undergo elastic scattering. The incoming `beam' proton wave function is then written in the form
\begin{equation}
|p\rangle~=~\sum a_i |\phi_i\rangle,
\label{eq:ai}
\end{equation}
and similarly for the incoming `target' proton. In terms of this $n$-channel eikonal model, the $pp$ elastic cross section has the form
\be
\frac{d\sigma_{\rm el}}{dt}~=~\frac{1}{4\pi}  \left| \int d^2b~e^{i\q_t \cdot \b} \sum_{i,k}|a_i|^2 |a_k|^2~(1-e^{-\Omega_{ik}(b)/2}) \right|^2,
\ee
where $-t=q_t^2$.

 We find that a 3-channel eikonal model is sufficient to describe all the 
elastic $pp$ data from 62.5 to 7000 GeV.  We assume that each diffractive eigenstate has the same weight, that is $a_i= 1/\sqrt{3}$ for $i=1,2,3$.
Moreover, we assume that the shape of the form factor which corresponds to the $\phi_i \to$Pomeron vertex has the form 
\begin{equation}
  V_i(t)=\gamma_i\beta_i(t),~~~~~~~{\rm where}~~~~~~~\beta_i(t)~=~e^{c_i t}/(1-t/d_i),
\label{eq:ff}
\end{equation}
and we take the square of the transverse size of each component proportional to its cross section. 
That is\footnote{Such an approach was used before in model B of \cite{KMRns} and in model (i) of \cite{KMRhardtosoft}.},
\begin{equation}
c_i~=~c\gamma_i, ~~~~~d_i=d/\gamma_i,
\label{eq:ad}
\end{equation}
so that each component has the same parton density at $b=0$, as would be expected from saturation  The different values of $\gamma_i$ of the eigenstates distort the original proton wave function (\ref{eq:ai}) after the interaction, leading to proton dissociation into some relatively low-mass proton resonances.  We choose the values of $\gamma_i$ so as to reproduce the cross section of low-mass dissociation\footnote{In this paper, $\sigma^{\rm SD}$ is the sum of the cross sections for `beam' and `target' dissociation.}, $\sigma^{\rm SD}_{{\rm low}M} \simeq 3$ mb, measured at the CERN-ISR \cite{lowM}; that is we take
\begin{equation}
\gamma_1=1.90,~~~~~\gamma_2=0.80,~~~~~\gamma_3=0.30.
\end{equation}

Given these diffractive eigenstates, we are left with a total of five free parameters to tune to describe the elastic $pp$ and $p\bar{p}$ data available in the 62.5 to 7000 GeV energy range.  First, we have the overall normalization, $\sigma_0$, which is driven by the proton-Pomeron coupling -- it is essentially the residue, $[\beta(0)]^2$, of the Pomeron pole up to some known normalization factor. Next we have two parameters, $c$ and $d$ of (\ref{eq:ad}), which specify the $t$-shapes of the various $\phi_i$-Pomeron couplings.  Then we have the intercept, $\alpha (0)\equiv 1+\Delta$, and slope, $\alpha'$, of the {\it effective} Pomeron trajectory. We call the trajectory `effective' since, at this stage, we account for only eikonal rescattering, and do not consider explicitly the enhanced diagrams, which account for the interactions between two or more Pomerons (such as described e.g. in Figs. 3 and 5 of Ref.\cite{KMRhardtosoft}).  However, the enhanced diagrams contribute implicitly, as we explain in a moment; it is known that their main effect is to `renormalize' the original
 Pomeron trajectory \cite{KMRns,KTP,ostap}.

\begin{figure} 
\begin{center}
\includegraphics[height=18cm]{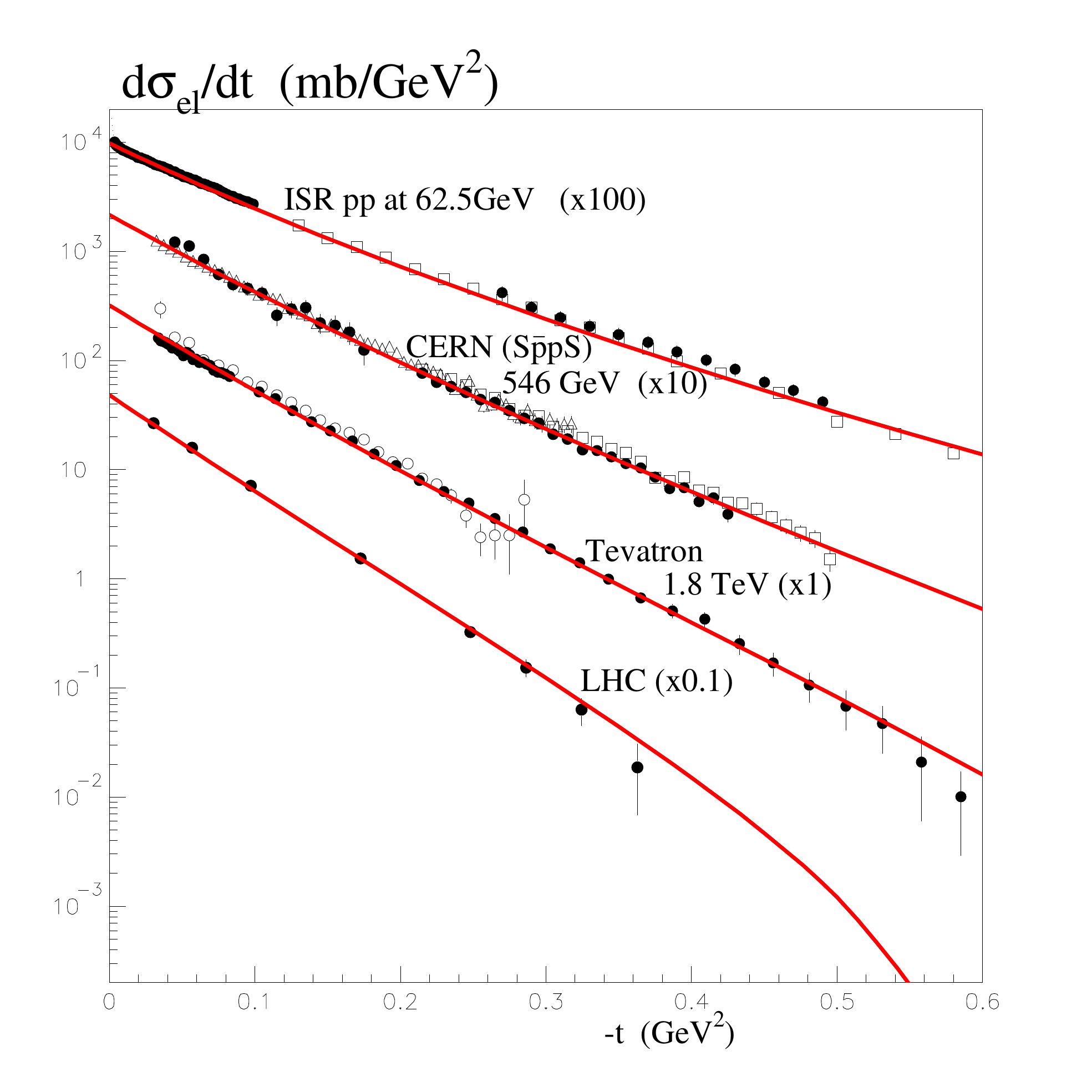}
\vspace{-0.8cm}
\caption{\sf The description of the data for the differential cross sections for $pp$ (or $p\bar{p}$) elastic scattering in the energy range 62.5 to 7000 GeV \cite{TOTEM2, SppS, Tev, ISR} using a 3-channel eikonal model. The Tevatron data with open and closed circles are those of the CDF and E710 collaborations respectively. Only very selected TOTEM points are shown, which have been read off their published plot. The excellent agreement of our model with the data for small $|t|$ is achieved with a very economical parametrization of the diffractive eigenstates. It is straightforward to describe the elastic data in the region of the diffractive LHC dip \cite{TOTEM1}, but at the expense of a more complicated parametrization of the form factors of the three eigenstates.}
\label{fig:elastic}
\end{center}
\end{figure}
Already this simple, physically motivated, eikonal model, with 
\begin{equation}
\sigma_0=22~{\rm mb},~~~~~c=0.4 ~\GeV^{-2},~~~~~ {\it d} =0.73 ~\GeV^2,~~~~~\Delta=0.14,~~~~~\alpha'=0.1~\GeV^{-2},
\end{equation}
provides a remarkably good description of all the collider elastic data for $|t| \lapproxeq 0.5~\GeV^2$, as can be seen from Fig~\ref{fig:elastic}.  The detailed structure of the cross sections at larger $|t|$, particularly in the region of the diffractive dip, depends sensitively on the fine structure of the form factors of the individual diffractive eigenstates. In general, there is enough freedom to describe the elastic data at higher $|t|$, at the expense of slightly more complicated forms of these form factors; which may have different structure for the different eigenstates.
Recall also that, at the moment, we have neglected the real part of the elastic amplitude,
which is essential only in the region of the diffractive dip.

It is not surprising that the value $\Delta=0.14$ found for the effective Pomeron is {\it larger} than the 0.08 (the value obtained when the amplitude was parametrized by one-pole-exchange without any multi-Pomeron corrections
 \cite{DLpl}), but is {\it smaller} than the intercept, $\Delta\sim 0.3$, expected for the bare Pomeron of the resummed NLL$(1/x)$ BFKL approach \cite{BFKL}.  In comparison with the simple model, we explicitly account for the non-enhanced eikonal absorption which suppresses the growth of the amplitude with energy.  Therefore to describe the same data we need a larger intercept ($\Delta=0.14$).  On the other hand, we do not explicitly include the enhanced diagrams, which would also slow down the growth of the cross section in the eikonal approach.  Thus we expect a smaller effective intercept than that given by BFKL.  Similar arguments apply to the slope of the effective trajectory, leading to a value ($\alpha'=0.1 ~\GeV^{-2}$)
 intermediate between the BFKL prediction\footnote{In perturbative QCD (and BFKL) the value of the slope, $\alpha'$, of the effective Pomeron trajectory is controlled by the transverse size of the incoming eigenstate. There is no other dimensionful parameter. Therefore it is natural to have a smaller $\alpha'$ when the Pomeron couples to a smaller-size diffractive eigenstate. Here, we choose $\alpha'_{ik}=\alpha'(1/i+1/k)/2$, where $i=1$ is the largest-size, and $i=3$ is the smallest-size, eigenstate}($\alpha' \gapproxeq 0$) and the old one-pole parametrization \cite{DL} ($\alpha'=0.25~ \GeV^{-2}$). Simultaneously we expect a renormalization of the effective pole residue, expressed in terms of $\beta_0$.

Some results of the 3-channel eikonal description of elastic data\footnote{
Recall that the pion loop insertion modifies the Pomeron trajectory at very small $t$ \cite{AG,KMR2000}. Indeed the presence of the $2\pi$ singularity at $t=4 m_{\pi}^2$ leads to some curvature in the $t$ behaviour of $d\sigma_{\rm el}/dt$. That is, to some variation of the local elastic slope $B(t)$. Including the pion loop gives an equally good description of the elastic data, but now the extrapolation to $t=0$ gives, via the optical theorem, a total cross of $\sigma_{\rm tot}=96.4$ mb.} are given in Table~\ref{tab:A2}.  Note that the value of the total cross section, $\sigma_{\rm tot}$, is close to that measured by the CDF collaboration. Nevertheless, our parametrization describes the E710 Tevatron $d\sigma_{\rm el}/dt$ data rather well, in spite of the fact that the E710 and CDF total cross section values differ by some 10$\%$.

\begin{table}[htb]
\begin{center}
\begin{tabular}{|c|c|c|c|c|c|}\hline
energy &   $\sigma_{\rm tot}$ &  $\sigma_{\rm el}$ &    $B$ &  $\sigma^{\rm SD}_{{\rm low}M}$ & $\sigma^{\rm DD}_{{\rm low}M}$ \\ \hline

  0.0625  &  43.8  &  7.3  &  13.4  &  3.0 & 0.3 \\
  0.546 &  65.2  & 13.4  &  16.1  &  4.8 & 0.5 \\
 1.8  &   79.3  &     17.9  &       18.0      &   5.9 &0.7 \\
 7   &   97.4    &    23.8    &     20.3  &      7.3 & 0.9 \\
14    &  107.5  &     27.2     &  21.6   &    8.1 & 1.1 \\
 100   &  138.8  &     38.1  &  25.8  &  10.4  & 1.6 \\
 \hline

\end{tabular}
\end{center}
\caption{\sf The results obtained from the 3-channel eikonal description of elastic (and quasi-elastic) $pp$ and $p\bar{p}$ data. $\sigma_{\rm tot}$,  $\sigma_{\rm el}$ and    $\sigma^{\rm SD,DD}_{{\rm low}M}$ are the total, elastic and low-mass single and double dissociation cross sections (in mb) respectively, where, in the latter cases, the mass of each dissociating system  satisfies $M<3$ GeV.  The cross section $\sigma^{\rm SD}$ is the sum of the dissociations of both the `beam' and `target' protons. $B$ is the mean elastic slope (in $\GeV^{-2}$), $d\sigma_{\rm el}/dt=e^{Bt}$, in the region $|t|<0.2~\GeV^2$.  The collider energies are given in TeV.}
\label{tab:A2}
\end{table}

Let us compare the results\footnote{Indeed, the analysis of the present paper may be regarded as a response to the `Lessons from the LHC' listed in Section 5 of \cite{KMRsantiago}, see also \cite{trento}.} shown in Table~\ref{tab:A2} with the inelastic cross section obtained by CMS, ATLAS and ALICE at 7 TeV. In this comparison we will take the observed cross sections at face value, with no attention paid to the (important) experimental errors, simply to illustrate the trends of the data. The measured value is defined as the cross section with at least two particles in some central (but far from complete) rapidity, $\eta$, interval.  For instance, ATLAS find 
$\sigma_{\rm inel}=60.3$ mb for the cross section of processes with $M>15.7$ GeV, that is $\xi =M^2/s >5 \times 10^{-6}$ \cite{ATLAS}.  After a model dependent extrapolation to cover the entire rapidity interval they obtain $\sigma_{\rm inel}=69.4$ mb.  CMS find a very similar result, namely 68.0 mb \cite{CMS1}. ALICE also get a similar result \cite{ALICE}.  
These estimates are about 5 mb lower than the recent TOTEM value \cite{TOTEM2}
\begin{equation}
\sigma_{\rm inel}=\sigma_{\rm tot}-\sigma_{\rm el}=73.5 \;{\rm mb}.
\end{equation}
The difference may be attributed to the extrapolated values being 5 mb deficient for low-mass diffraction. (The extrapolation in the high-mass interval is confirmed by the ATLAS measurement $d\sigma/d\Delta\eta \simeq d\sigma/d{\rm ln} M^2 \simeq 1$ mb per unit of rapidity \cite{Atlas}.) More specifically, if we define low mass to be $M<3$ GeV, then, noting that the unmeasured interval from $M=15.7$ to $M=3$ GeV gives $\Delta{\rm ln} M^2 =3.3  $, it follows that the ATLAS, CMS results imply $\sigma_{\rm inel}^{{\rm high}M}\simeq 64$ mb. Then using the TOTEM result we find that low-mass diffractive dissociation is expected to have a cross section
\begin{equation}
\sigma_{\rm inel}^{{\rm low}M}\simeq 73.5-64 ~=~ 9.5 \;{\rm mb}.
\label{eq:low}
\end{equation}

Coming back to Table~\ref{tab:A2}, we see that at 7 TeV
\begin{equation}
\sigma_{\rm inel}~=~\sigma_{\rm tot}-\sigma_{\rm el}~=~73.6~{\rm mb},
\end{equation}
in agreement with the TOTEM measurement. Moreover, we indeed have a rather large cross section for low-mass diffractive dissociation
\begin{equation}
\sigma^{\rm SD+DD}_{{\rm low}M}~=~7.3+0.9~=~8.2 ~{\rm mb},
\end{equation} 
in satisfactory agreement with (\ref{eq:low}).

\section{Diffractive dissociation  \label{sec:diff}}

Now that we have a good parametrization of the opacity of the proton, we can study the dynamics of proton dissociation accounting for the gap survival effect. In our formalism there are two distinct ways in which dissociation can occur.

First, as we have discussed, there is dissociation into {\it low-mass} systems. This is described by the proton excitations which occur in the diffractive eigenstate approach. After rescattering the original proton wave function, (\ref{eq:ai}), is distorted and is decomposed into the various hadronic eigenstates.  The cross section of the low-mass proton excitations are given in the last two columns of Table \ref{tab:A2}. Since we allow for only three eigenstates, we cannot account for high-mass dissociation in this way.

\begin{figure} 
\begin{center}
\includegraphics[height=3cm]{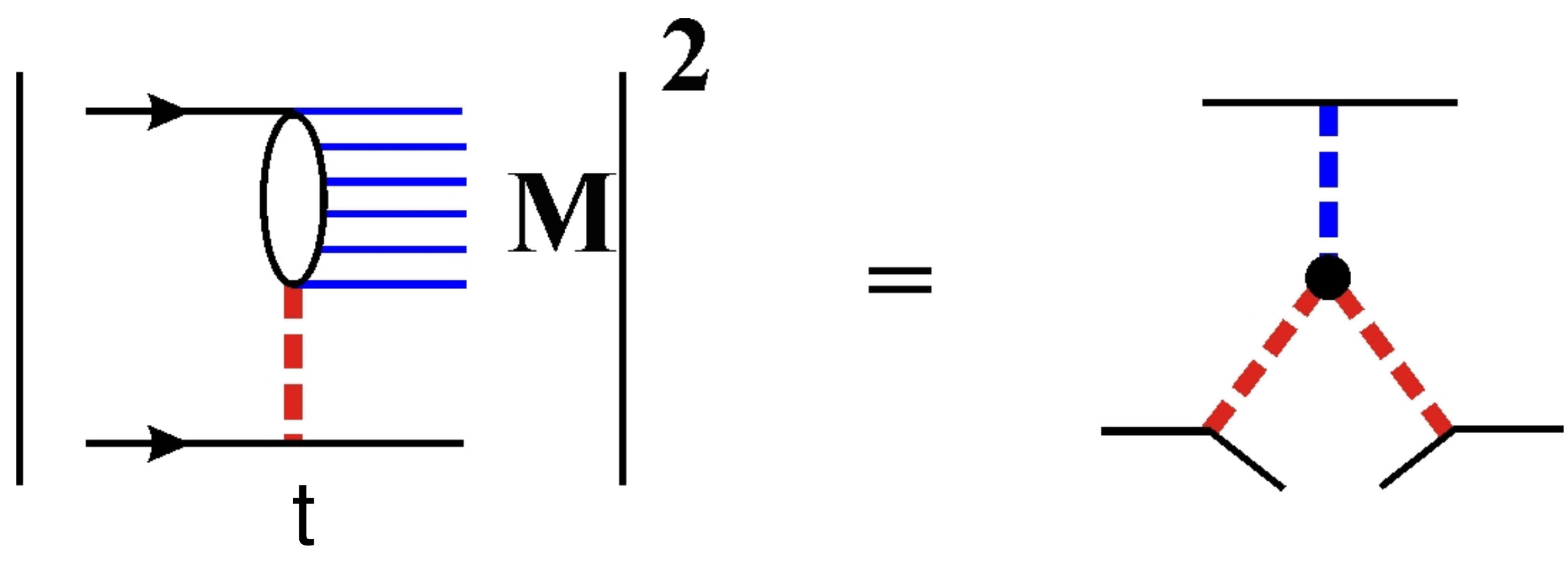}
\caption{\sf The triple-Pomeron diagram which describes the process $pp \to X+p$ where a proton dissociates into a system of high mass $M$.  Neglecting absorptive effects, the cross section is given by (\ref{eq:3P}).}
\label{fig:PPP}
\end{center}
\end{figure}
The process $pp \to X+p$, where one proton dissociates into a system $X$ of 
{\it high-mass} $M$ is conventionally studied in terms of the triple-Pomeron diagram of Fig. \ref{fig:PPP}. 
In the absence of absorptive corrections, the
corresponding cross section is given by
\be
\frac{M^2 d\sigma}{dtdM^2}~=~g_{3P}(t)\beta(0)\beta^2(t)~\left(\frac{s}{M^2}\right)^{2\alpha(t)-2}~\left(\frac{M^2}{s_0}\right)^{\alpha(0)-1},
\label{eq:3P}
\ee
where $\beta(t)$ is the coupling of the Pomeron to the proton and $g_{3P}(t)$ is the triple-Pomeron coupling.
The coupling $g_{3P}$ is obtained from a fit to lower energy data.  Mainly it is the data on proton dissociation taken at the CERN-ISR with energies from $23.5 \to 62.5$ GeV.

\subsection{Absorptive effects in high-mass diffractive dissociation}

The problem, in the above determination of $g_{3P}$, is that this is an effective vertex with coupling
\be
g_{\rm eff}~=~g_{3P}*S^2
\ee
which already includes the suppression $S^2$ -- the probability that no other secondaries, simultaneously produced in the same $pp$ interaction, populate the rapidity gap region.  Recall that the survival factor $S^2$ depends on the energy of the collider.  Since the opacity $\Omega$ increases with energy, the number of multiple interactions, $N \propto \Omega$, grows\footnote{This is because at larger optical density $\Omega$ we have a larger probability of interactions.}, leading to a smaller $S^2$.  Thus, we have to expect that the naive triple-Pomeron formula with the coupling \cite{abk,KKPT}, 
measured at relatively low collider energies will appreciably overestimate the cross section for high-mass dissociation at the LHC. A more precise analysis \cite{luna} accounts for the survival effect $S^2_{\rm eik}$ caused by the eikonal rescattering of the fast `beam' and `target' partons.  In this way, a coupling $g_{3P}$ about a factor of 3 larger than $g_{\rm eff}$ is obtained, namely $g_{3P} \simeq 0.2g_N$, where $g_N$ is the coupling of the Pomeron to the proton. The analysis of Ref. \cite{luna} enables us to better take account of the energy dependence of $S^2_{\rm eik}$.
We therefore use this formulation to calculate the cross section $d\sigma/d\Delta\eta$ of  Fig. \ref{fig:Atlas}. Details of the calculation are given in the Appendix.

  At this stage, our prediction of the cross section still overestimates the ATLAS data with $\Delta\eta >5$ of Fig. \ref{fig:Atlas}; but there is still one further absorptive effect that we must include.
\begin{figure} 
\begin{center}
\includegraphics[height=4cm]{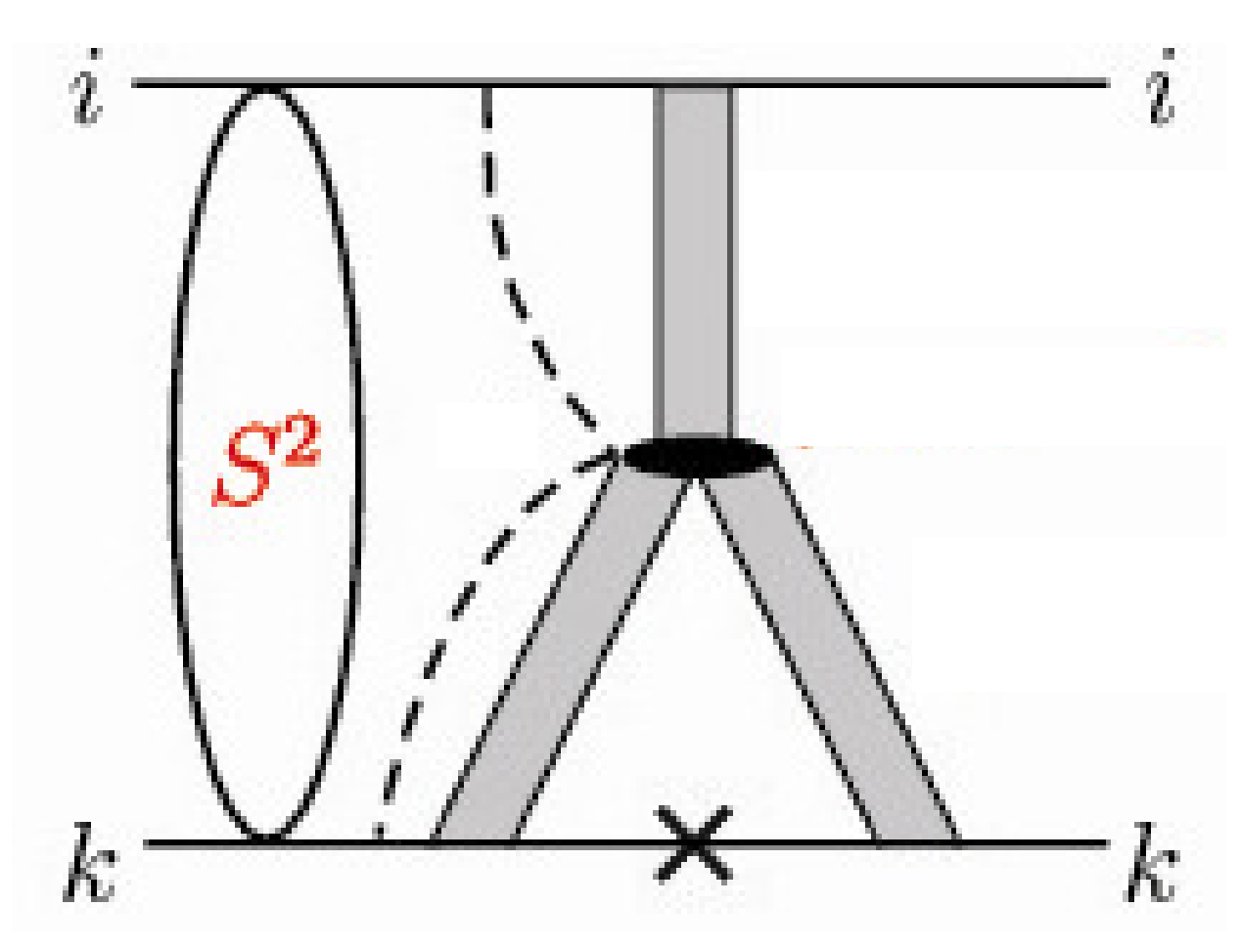}
\caption{\sf A sketch of the diagram that describes single proton diffractive dissociation, $pp \to X+p$.  System $X$ has mass $M$. Each of the three shaded areas represents an effective Pomeron as determined by the 3-channel eikonal analysis in Section \ref{sec:elastic} with a renormalized trajectory which implicitly includes enhanced rescattering. The dashed lines represent the rescattering of the partons forming the triple-Pomeron vertex with the incoming diffractive eigenstates of one or the other protons. }
\label{fig:14}
\end{center}
\end{figure}
Recall that in our 3-channel eikonal determination of the opacity from the elastic data, we used an effective Pomeron with a renormalized trajectory, with parameters $\Delta$ and $\alpha'$. In this way we already implicitly include the corrections to the Pomeron trajectory caused by rescattering due to the enhanced diagrams. However, it does not account for the renormalization of the triple-Pomeron vertex due to the rescattering of the partons which form the vertex with one or the other incoming protons \cite{KMRns}. This rescattering is indicated by the dashed lines in Fig. \ref{fig:14}. This absorptive effect can be calculated explicitly in an analogous way to the renormalization of the proton-Pomeron coupling as was done in the eikonal amplitude of eq.(\ref{eq:1}). The summation of the rescatterings due to these non-enhanced\footnote{The rapidity of the triple-Pomeron vertex in these rescatterings is fixed by the value of $M^2$, and so we have no enhancement arising from the integration over the large rapidity interval allowed for the vertex position in the enhanced case.} (eikonal) interactions again leads to the factor exp$(-\Omega/2)$, analogous to that in (\ref{eq:1}).  Recall that here we need to use the full $S$-matrix, $S=1+iT=e^{-\Omega/2}$.  Moreover, to account for the rescattering on both of the incoming protons, we must include the absorptive factor\footnote{This is similar to the approach used in \cite{KP}.}
\be
{\rm exp}(-(\Omega_i+\Omega_k)/2),
\label{eq:last}
\ee
where the indices $i,k$ refer to particular diffractive eigenstates in the beam and target proton respectively. These must be summed over. A more detailed discussion of this absorptive factor can be found in Ref. \cite{KMRns}\footnote{In the triple-Regge analysis of \cite{luna} the suppression due to (\ref{eq:last}) was not evaluated explicitly, but included in the effective value $\lambda_{\rm eff}=g_{3P}/g_N\simeq0.2$. We have therefore used the larger value, $\lambda=0.3$, which was obtained in model B3, with a 3-channel eikonal, of \cite{KMRns}; and which, in the present formalism, provides a good description of the Tevatron data in the triple-Regge region.}.
Including the factor (\ref{eq:last}), we obtain the rapidity gap cross section shown by the curve for $\Delta\eta >5$ in Fig. {\ref{fig:Atlas}.  We see that the curve is in general agreement with the ATLAS measurements. We discuss below the comparison of the curve with the data, first, in terms of normalization, and then in terms of shape.

\subsection{Normalization of $d\sigma/d\Delta\eta$}

Before we proceed, we need to specify exactly what ATLAS has measured and exactly what we have calculated.
In the ATLAS measurement \cite{Atlas} the large rapidity gap events
are selected using the combined information from the inner detector tracks and the
calorimeters
detecting particles in the rapidity interval $|\eta|<4.9$, while the larger rapidity interval up to proton $y=\pm 8.9$ is uninstrumented.  ATLAS measure $d\sigma/d\Delta\eta$ with $\Delta\eta$ defined by the larger of the two empty $\eta$ regions extending between the edges of the detector acceptance at $\eta=4.9$ or $\eta=-4.9$ and the nearest track or cluster, passing the selection requirements, at smaller $|\eta|$. The gap size measured by ATLAS relative 
to $\eta=\pm 4.9$ lies in the range $0<\Delta\eta<8$, such that, for example, $\Delta\eta=8$ implies that
no final state particles are produced above a transverse momentum threshold
$p_T^{\rm cut}=200$ MeV in one of the regions $-4.9<\eta<3.1$ or $-3.1<\eta<4.9$.

We calculated $d\sigma/d\Delta\eta'$ with $\Delta\eta'=\Delta\eta+4$ using the triple-Pomeron formula
(\ref{eq:3P}),
but allowing for the large `absorptive' suppression factors, arising from the probability that the rapidity gap survives eikonal and enhanced rescattering.  The procedure is summarised in the Appendix.

The ATLAS $d\sigma/d\Delta\eta$ cross section measurements shown in Fig. {\ref{fig:Atlas} may include both single proton dissociation and double dissociation.  Part of the double dissociation, where the masses satisfy $M<3$ GeV, is already included in our computation via the 3-channel eikonal formalism.  Nevertheless, high-mass double dissociation, $3<M<7$ GeV\footnote{$M=7$ GeV is the minimal mass which, according to Monte Carlo simulations, can be reconstructed in the ATLAS experiment\cite{Atlas}.}, is missed, indicating that we should slightly underestimate the ATLAS data.

In summary, we see that the ATLAS data indicate the important role of absorptive effects, described by the survival factors $S^2$.  Moreover, it is not sufficient to consider only eikonal rescattering of the fast protons (which are, more or less, included in the present Monte Carlos as the multiple interaction option (MI). It is necessary to allow for absorption arising from enhanced diagrams, in which we account for the vertices with a large number of Pomerons.

\subsection{Shape of the $\Delta\eta$ distribution   \label{sec:shape}}

From the naive triple-Pomeron formula, (\ref{eq:3P}), and noting $\xi=M^2/s$ and $\Delta\eta'\simeq -{\rm ln}\xi$, we expect
\be
\frac{d\sigma}{d\Delta\eta'}~~\propto~~e^{(\Delta\eta')\Delta}~~~~~~~~~~~~ {\rm with}~~~~~ \Delta=0.14.
\label{eq:D}
\ee
 How is this result modified by a more careful treatment, including the influence of the gap survival factors? There are three effects.

First, we note that the cross section measured by ATLAS, Fig. {\ref{fig:Atlas}, has already been integrated over the transverse momentum. Thus the quantity which controls the $\Delta\eta'$ behaviour in (\ref{eq:D}), is not simply $\Delta$, but rather is $\Delta-2\alpha'\langle |t| \rangle$. Preliminary TOTEM data \cite{TOTEMSD} indicate that the slope $B$ of the cross section for the single diffractive dissociation is about 10 $\GeV^{-2}$, that is $\langle |t| \rangle \simeq 0.1 ~\GeV^2$. In our analysis with $\alpha' \simeq 0.1 ~\GeV^{-2}$, this would shift the value of $\Delta$ by $-0.02$. 

To see the next effect we must return to Fig. \ref{fig:14} (or Fig. \ref{fig:3Rb}). In impact parameter representation, the single diffractive dissociation cross section, described by the triple-Pomeron diagram, can be regarded as a convolution of elastic cross sections, $\sigma_{\rm down}$ (in the lower part of Fig. \ref{fig:14}) and inelastic cross section (in the upper part of the diagram) \cite{KMRns}.  The elastic contribution is, \begin{equation}
\sigma_{\rm down} ~\propto~|T_{\rm down}(b)|^2~\propto {\rm exp}(-b^2/2B_{\rm down}),
\end{equation}
while for the inelastic contribution, the amplitude for the upper part of the diagram has the form $|T_{\rm up}|~\propto~{\rm exp}^{-b^2/4B_{\rm up}}$. The slopes,
\begin{equation}
B_{\rm up}({\rm inel})~=~B_0({\rm up)}+\alpha'{\rm ln}(\xi s),~~~~~~~~~B_{\rm down}({\rm el})~=~B_0({\rm down)}-\alpha'{\rm ln}\xi, 
\end{equation}
depend on the rapidity of the vertex, that is on $\Delta\eta' \simeq -{\rm ln}\xi$.  Thus for larger $\Delta\eta$, we enlarge the slope coming from the elastic part and decrease the slope coming from the inelastic part.  
 Since the elastic cross section is described by the amplitude 
 squared, that is concentrated at smaller $b$, the typical 
value of $b$ becomes smaller, where $S^2$ is lower (see also Fig. 2 of~\cite{ostap2}).  This additional absorptive effect decreases the value of $\Delta$ by about $0.04$.  

Finally, we have an analogous effect coming from the absorption in the triple-Pomeron vertex, which again decreases the value of $\Delta$ by about $0.04$. 

Thus the absorptive effects make the $\Delta\eta$ distribution much flatter than that which might be expected based on original triple-Pomeron expression (\ref{eq:3P}). First, the enhanced absorptive corrections decrease the effective Pomeron intercept $\Delta$ in comparison with that of the bare (unscreened) Pomeron.
 Next, the lower survival probability of gaps of larger size 
 further suppresses the increase of $d\sigma/d\Delta\eta'$
with increasing $\Delta\eta'$.  Moreover, at large LHC energies, the majority of secondaries comes from minijet fragmentation, and the transverse momenta of these minijets ($p_{t}>k_{\rm min}\sim 2$ GeV) are not small. Thus in the very large $\Delta\eta$ region, corresponding to low $M$, the cross section $d\sigma/d\Delta\eta'$ starts to decrease, as the
available phase space becomes insufficient to generate relatively large $p_t$ minijets (see Fig. 9 of \cite{KMRhardtosoft}).

The fact that our curve plotted in Fig. \ref{fig:Atlas} is flatter than the behaviour of the 
ATLAS data, may indicate that the absorptive effects are too strong in the present simplified model, which does not account for the parton transverse momenta distribution. Indeed,  partons with larger transverse momentum have smaller absorptive cross sections. Therefore in the model of \cite{KMRhardtosoft}, where the $p_t$ dependence was accounted for, we obtained a steeper behaviour
(see Fig. 9 of~\cite{KMRhardtosoft}).
 Thus we need a more complicated and more complete model for a detailed description of the shape of the $\Delta\eta$ distribution of the LHC data.

\section{Conclusions}

Although there are a lot of experimental measurements of high energy $pp$ and $p\bar{p}$ elastic scattering, we have comparatively little experimental information on `soft' diffractive dissociation processes. For the latter processes, most of the information has come from measurements at the CERN-ISR. Indeed, the CERN-ISR experiments accumulated data in the triple-Regge domain, which allowed the triple-Pomeron coupling to be extracted, provided secondary Reggeons were included, and provided, as we have seen in Section \ref{sec:diff}, careful account is taken of absorptive corrections. Moreover, the CERN-ISR has, to date, made the only estimate of low-mass diffractive dissociation - a measurement invaluable in constraining the diffractive eigenstates.  The CDF collaboration have added some more information in the triple-Regge region.   This sums up the pre-LHC information on soft diffraction.

The LHC is already starting to shed new light on soft diffractive processes. Besides the TOTEM data on elastic scattering, we have the ATLAS data of Fig. \ref{fig:Atlas} on cross sections with rapidity gaps.  In this paper, we have shown that the TOTEM data, together with other elastic data from low energy colliders, can be described naturally in a conventional 3-channel eikonal model, where the opacity $\Omega$ is driven by {\it single} Pomeron exchange.  There is no need to introduce a second Pomeron with a higher intercept to reproduce the LHC data measured by TOTEM at 7 GeV.

We then used the proton opacity to calculate the cross section $d\sigma/d\Delta\eta$  for rapidity gaps $\Delta\eta>5$, in order to compare with the ATLAS measurements.  The opacity allows us to determine the absorptive effects which reduce the value of the cross section, and which change its $\Delta\eta$ behaviour. Within this model, this is an essentially parameter-free calculation, since we use the known triple-Pomeron coupling that had been tuned to describe the lower energy CERN-ISR and Tevatron data. The cross section at larger $\Delta\eta$ comes from smaller $b$, where the absorption is stronger, and slows down the growth of $d\sigma/d\Delta\eta$ with increasing $\Delta\eta$.  For this reason the Pomeron intercept extracted from the $d\sigma/d\Delta\eta$ data, using the naive triple-Pomeron formula, is smaller than that of the original bare Pomeron pole. 

The data and the parameter-free calculation were compared in Fig. \ref{fig:Atlas}. We would not have expected our prediction to be so 
 good, bearing in mind the complexity of the coherence effects 
that underlie the nature of diffraction. The theoretical $\Delta\eta$ dependence in Fig. \ref{fig:Atlas} is a bit flatter than is observed. However, note that we have used a pure `soft' framework; that is, we assume that the transverse momenta of the partons, that form the Pomeron, are limited. A model which allows for the higher transverse momenta was presented in \cite{KMRhardtosoft}. It was found that this, indeed, gives a steeper $\Delta\eta$ behaviour (see Fig. 9 of \cite{KMRhardtosoft}), as required by the ATLAS data.

So to summarize, the main message is that it is possible to describe all the elastic $pp$ (and $p\bar{p})$ collider data for $|t| \lapproxeq 0.4~\GeV^2$ in terms of a 3-channel eikonal model of a {\it single} Pomeron, and, secondly, that absorptive corrections appreciably modify the value and the $\Delta\eta$ bevaviour of diffractive cross sections with rapidity gaps.

\section*{Appendix}

\begin{figure} 
\begin{center}
\includegraphics[height=4cm]{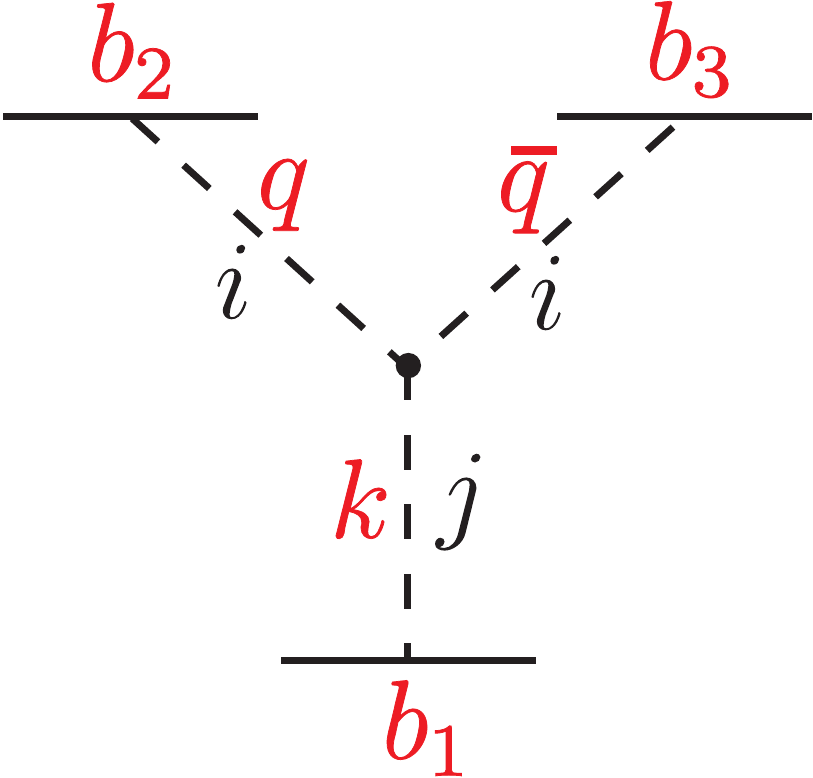}
\caption{\sf A schematic diagram showing the notation of the impact parameters arising in the calculation of the screening corrections to the $iij$ triple-Pomeron diagram. The conjugate momenta to $b_1,b_2,b_3$ are $k_t,q_t,\bar{q}_t$. If $k_t=0$, then $\bar{q}_t=q_t$.}
\label{fig:3Rb}
\end{center}
\end{figure}
Here we outline how we calculate the absorptive corrections to the triple-Pomeron formula, (\ref{eq:3P}).
We follow the procedure of Ref. \cite{luna}. We first take Fourier transforms with respect to the impact parameters specified in Fig. \ref{fig:3Rb}. We then obtain\footnote{Note that $e^{i\vec{k}_t \cdot \vec{b}_1}=1$ as $k_t=0$.}
\be
\frac{M^2 d\sigma}{dtdM^2}~=~A\int\frac{d^2b_2}{2\pi}e^{i\vec{q}_t \cdot \vec{b}_2} F_i(b_2)\int\frac{d^2b_3}{2\pi}e^{i\vec{q}_t \cdot \vec{b}_3} F_i(b_3)\int\frac{d^2b_1}{2\pi} F_j(b_1),
\label{eq:3Rb}
\ee
where
\be
F_i(b_2)~=~\frac{1}{2\pi \beta_i(q_t=0)}\int d^2q_t \beta_i(q_t)\left(\frac{s}{M^2}\right)^{-\alpha^\prime_i q^2_t} e^{b^\prime_{iij}q^2}e^{i\vec{q}_t \cdot \vec{b}_2},
\label{eq:Fi}
\ee
\be
F_j(b_1)~=~\frac{1}{2\pi \beta_j(k_t=0)}\int d^2k_t \beta_j(k_t)\left(\frac{M^2}{s_0}\right)^{-\alpha^\prime_j k^2_t} e^{-b^\prime_{iij}k^2_t},
\label{eq:Fj}
\ee
and where 
\be
A~=~\beta_j(0)\beta_i^2(0)g_{iij}(0)\left(\frac{s}{M^2}\right)^{2\alpha_i(t)-2}\left(\frac{M^2}{s_0}\right)^{\alpha_j(0)-1}.
\label{eq:A}
\ee
Here, we have assumed
\be
g_{3P}(t)~=~g_{3P}(0)~{\rm exp}(b'(q^2_t+\bar{q}^2_t)).
\ee

After integrating (\ref{eq:3Rb}) over $t$, the cross section becomes
\be
\frac{M^2 d\sigma}{dM^2}~=~A\int\frac{d^2b_2}{\pi}\int\frac{d^2b_1}{2\pi} |F_i(b_2)|^2 F_j(b_1) \cdot S^2(\vec{b}_2-\vec{b}_1),
\label{eq:result}
\ee
where here we have included the screening correction $S^2$, which depends on the separation in impact parameter space, $(\vec{b}_2-\vec{b}_1)$, of the incoming protons
\be
S^2(\vec{b}_2-\vec{b}_1)~\equiv~{\rm exp}(-\Omega(\vec{b}_2-\vec{b}_1))
\ee
The result (\ref{eq:result}) has been written for a single channel eikonal, but
the generalization of the formalism to a 3-channel eikonal is straightforward, see \cite{luna}.

\section*{Acknowledgements}
We thank Tim Martin, Paul Newman and Andy Pilkington for discussions. MGR thanks the IPPP at the University of Durham for hospitality. This work was supported by the grant RFBR 11-02-00120-a
and by the Federal Program of the Russian State RSGSS-65751.2010.2.


\thebibliography{}


\bibitem{TOTEM1} TOTEM Collaboration, Europhys. Lett. {\bf 95}, 41001 (2011).

\bibitem{TOTEM2} TOTEM Collaboration, Europhys. Lett. {\bf 96}, 21002 (2011).

\bibitem{Atlas} ATLAS Collaboration, arXiv{1201.2808}

\bibitem{abk} A.B.~Kaidalov,
  Phys.\ Rept.\  {\bf 50}, 157 (1979).

\bibitem{Amaldi} U.~Amaldi, {\it  Elastic and inelastic processes at
 the Intersecting Storage Rings --
the experiments and their impact parameter description}: Erice 1973, Proceedings, Laws of Hadronic Matter, New York 1975, 673-741;\\ 
 U.~Amaldi, M.~Jacob and G.~Matthiae,
  Ann.\ Rev.\ Nucl.\ Part.\ Sci.\  {\bf 26}, 385 (1976).

\bibitem{SppS} UA4 Collaboration, Phys. Lett. {\bf B147}, 385 (1984);\\
UA4/2 Collaboration, Phys. Lett. {\bf B316}, 448 (1993); \\
UA1 Collaboration, Phys. Lett. {\bf B128}, 336 (1982).


\bibitem{Tev} E710 Collaboration, Phys. Lett. {\bf B247}, 127 (1990);\\
CDF Collaboration, Phys. Rev. {\bf D50}, 5518 (1994).

\bibitem{DL2} A. Donnachie and P.V. Landshoff, arXiv:1112.2485.

\bibitem{GW} M.L. Good and W.D. Walker, Phys. Rev. {\bf 120}, 1857 (1960).

\bibitem{KKMRZ} V.A. Khoze et al., Eur. Phys. J. {\bf C69}, 85 (2010).

\bibitem{KMRns}  M.G. Ryskin, A.D.Martin and V.A. Khoze, Eur. Phys. J. {\bf C54}, 199 (2008).

\bibitem{KMRhardtosoft} M.G. Ryskin, A.D.Martin and V.A. Khoze, Eur. Phys. J. {\bf C71}, 1617 (2011).

\bibitem{lowM} L.~Baksay {\it et al.}, Phys.\ Lett.\ {\bf B53}, 484 (1975); \\
R.~Webb {\it et al.}, Phys.\ Lett.\ {\bf B55}, 331 (1975); \\
L.~Baksay {\it et al.}, Phys.\ Lett.\ {\bf B61}, 405 (1976); \\
H.~de Kerret {\it et al.}, Phys.\ Lett.\ {\bf B63}, 477 (1976); \\
G.C.~Mantovani {\it et al.}, Phys.\ Lett.\ {\bf B64}, 471 (1976).


\bibitem{KTP} A.B. Kaidalov, L.A. Ponomarev and K.A. Ter-Martirosyan, Sov. J. Nucl.     
Phys. {\bf 44}, 468 (1986).

\bibitem{ostap} S.~Ostapchenko,
  Phys.\ Rev.\   {\bf D81}, 114028 (2010).

\bibitem{ISR} N. Kwak et al., Phys. Lett. {\bf B58}, 233 (1975);\\
U. Amaldi et al., Phys. Lett. {\bf B66}, 390 (1977);\\
L. Baksay et al., Nucl. Phys. {\bf B141}, 1 (1978).

\bibitem{DLpl} A.~Donnachie and P.V.~Landshoff,  Phys. Lett. {\bf B296}, 227 (1992).

\bibitem{BFKL} M.~Ciafaloni, D.~Colferai and G.~Salam, Phys. Rev. {\bf D60}, 114036 (1999); \\
G.~Salam, JHEP {\bf 9807}, 019 (1998); Act. Phys. Pol. {\bf B30}, 3679 (1999);\\
V.A.~Khoze, A.D.~Martin, M.G.~Ryskin and W.J. Stirling, Phys. Rev. {\bf D70}, 074013 (2004).

\bibitem{DL} A.~Donnachie and P.V.~Landshoff,  Nucl.\ Phys.\ {\bf B231}, 189 (1984).

\bibitem{AG} A.A. Anselm and V.N. Gribov, Phys. Lett. {\bf B40}, 487 (1972).

\bibitem{KMR2000} V.A. Khoze, A.D. Martin and M.G. Ryskin, Eur. Phys. J. {\bf C18}, 167 (2000).

\bibitem{KMRsantiago} A.D. Martin, M.G. Ryskin and V.A. Khoze, arXiv:1110.1973.
\bibitem{trento} A.~D.~Martin, V.~A.~Khoze and M.~G.~Ryskin,
  arXiv:1202.4966 [hep-ph].

\bibitem{ATLAS} ATLAS Collaboration,  Nature Commun.\  {\bf 2}, 463 (2011)
[arXiv:1104.0326].

\bibitem{CMS1} CMS Collaboration, Note CMS-PAS-FWD-11-001, (2011).

\bibitem{ALICE} M.G.~Poghosyan, for the ALICE Collaboration,
  arXiv:1109.4510 [hep-ex]. 


\bibitem{KKPT}A.B.~Kaidalov, V.A.~Khoze, Y.F.~Pirogov and N.L.~Ter-Isaakyan,
  Phys.\ Lett. {\bf B45}, 493 (1973).

\bibitem{luna} E.G.S. Luna, V.A. Khoze, A.D. Martin and M.G. Ryskin, Eur. Phys. J. {\bf C59}, 1 (2009).

\bibitem{KP} A.~B.~Kaidalov and M.~G.~Poghosyan,
  arXiv:0909.5156 [hep-ph].

\bibitem{TOTEMSD} M.~Deile [on behalf of the TOTEM collaboration],
talk at the 14th Workshop on Elastic and Diffractive Scattering
(EDS Blois Workshop),
December 15-21, 2011,
Qui Nhon, Vietnam.
 
\bibitem{ostap2}  S.~Ostapchenko,
  Phys.\ Lett.\ B {\bf 703}, 588 (2011)
  [arXiv:1103.5684 [hep-ph]].
\end{document}